\documentclass[12pt,fleqn,cite,epsfig]{article}
\usepackage{cite}
\usepackage[english]{babel}
\usepackage[utf8]{inputenc}
\usepackage{graphicx}
\usepackage{longtable}
\usepackage{tabularx}
\usepackage{amsmath}
\usepackage{amsfonts}
\usepackage{amssymb}
\usepackage[paper=a4paper,
            marginparwidth=0in,     
            marginparsep=.05in,       
            margin=.65in,               
            includemp]{geometry}
\usepackage{hyperref}
\newcommand{\cP}{{\mathcal P}}
\newcommand{\C}{{\mathcal C}}
\newcommand{\bse}{\begin{subequations}}
\newcommand{\ese}{\end{subequations}}
\numberwithin{equation}{section}

\title{\textbf{F-string Solution in $AdS_4 \times \C\cP^3$ PP-wave Background}}

\author{Gourav Banerjee\footnote{$gourav\_9124@ap.ism.ac.in$},
 Binata Panda \footnote{panda.b.ap@ismdhanbad.ac.in}\\
 \textit{Department of Applied Physics,}\\
 \textit{ Indian Institute of Technology (ISM), Dhanbad 826004, India}}

\date{}

\begin{document}
\maketitle
\begin{abstract}
We present supergravity solution for F-string in pp-wave background 
obtained from ${AdS_4\times \C\cP^3}$ with zero flat directions.The classical solution is shown to break all
space-time supersymmetries. We explicitly write down the standard as well as supernumerary Killing spinors both for the 
background and F-string solution.

\end{abstract}

\section{Introduction}
Study of string theory in plane wave  background \cite{penrose, Horowitz:1994rf, Gueven:2000ru,Metsaev:1999gz,Metsaev:2001bj,Metsaev:2002re,Blau:2001ne,Blau:2002dy,Blau:2002mw,
Berenstein:2002jq,mukhi,GO,tseytlin,sonn,mohsen,chu,lee,Bala,Taka}
has drawn special attention, in  the context of establishing
AdS/CFT like dualities. In several cases, pp-waves are also  known to be maximally supersymmetric 
solutions of supergravities in 
various dimensions~\cite{meessen,pope,hull}. Moreover, they are known to  
provide exactly solvable string theories from worldsheet 
point of view. In this connection, many  classes of D-brane solutions are investigated in different 
pp-wave background geometry \cite{Biswas:2002yz,Kumar:2002ps,Alishahiha:2002zu,Hassan:2003ec,Alishahiha:2002rw}.

In this paper we are interested in a Penrose limit of the ${AdS_4\times \C\cP^3}$ background
considered in \cite{Nishioka:2008gz}. A good understanding of Type IIA string theory on ${AdS_4\times \C\cP^3}$ is of
great importance, due to its appearance in another example of $AdS/CFT$ duality i.e. $ AdS_4/CFT_3$
\cite{Maldacena:1997re,Witten:1998qj,Gubser:1998bc}. The new 
duality is motivated by the conjecture that  ${\cal{N}}=8$
Superconformal Chern-Simons theories in three dimensions describe dynamics of
multiple M2-branes~\cite{Schwarz:2004yj,Bagger:2006sk,Bagger:2007jr,Gustavsson:2007vu,Gomis:2008be,Bagger:2007vi,Gaiotto:2007xh,Gaiotto:2008sd,Bandres:2008vf,VanRaamsdonk:2008ft,Lambert:2008et,Benvenuti:2008bt,Krishnan:2008zm,Bandres:2008kj,Bagger:2008se}. The ABJM model~\cite{Aharony:2008ug} describes
a new example of the  $AdS/CFT$ duality involving
${\cal N}=6$ superconformal ${\rm SU}(N)\times
{\rm SU}(N)$ Chern-Simons theory in three dimensions and
M-theory on $ AdS_4\times  S^7/{\mathbb Z}_k$, where $k$
is the level of the Chern-Simons action. At strong coupling, the M-theory on $ AdS_4\times  S^7/{\mathbb Z}_k$,
can be effectively described by  type IIA
superstring theory on the $ AdS_4\times \C\cP^3$ background~\cite{Aharony:2008ug}.  Construction of
string and brane actions for IIA strings in $AdS_4\times \C\cP^3$  has been
discussed extensively in \cite{Arutyunov:2008if,Stefanski:2008ik,Bonelli:2008us,D'Auria:2008cw,Sorokin:1985ap,Gomis:2008jt,Grassi:2009yj,Cagnazzo:2009zh}. In the aim to consider CFT's with a defect, $AdS/dCFT$ duality, 
both supersymmetric and non-supersymmetric embeddings
of $D$-branes in $AdS_4\times \C\cP^3$ have been described in \cite{Chandrasekhar:2009ey}.
In this context, pp-wave geometries arising from the $AdS_4\times \C\cP^3$ has also been studied. Pp-wave backgrounds 
obtained 
by taking Penrose limit of ${AdS_4\times \C\cP^3}$ with zero space like isometry~\cite{Nishioka:2008gz}, with one and 
two
space like isometries \cite{Grignani:2009ny} were studied. The case of a D2-brane in a general pp-wave background 
obtained from ${AdS_4\times \C\cP^3}$ was considered in \cite{AliAkbari:2010rs} with electric and magnetic fields
turned on.

In view of these developments, herewith we present the classical solution for
fundamental string in pp-wave background 
obtained from ${AdS_4\times \C\cP^3}$ with zero flat directions considered in \cite{Nishioka:2008gz}. We explicitly
verify that our solutions satisfy the field equations. The 
supersymmetric properties of the solution  is discussed by explicitly finding the standard as well as 
supernumerary Killing spinors.  The rest 
of the paper is organised as follows.  In section-\ref{pp wave background}, we review the pp-wave backgrounds with 
different number of space like isometries. In section-\ref{supergravity solution}, we present the F-string solution
in the  ${AdS_4\times \C\cP^3}$ pp-wave background. The supersymmetry analysis is discussed
in section-\ref{supersymmtry analysis}. We analyze the supersymmery of the background and the F-string solution in
this background by solving the Killing spinor equations. Section-\ref{conclusion}
contains our conclusions and discussions. In the appendix-\ref{appendix}, we have listed the non-vanishing Christoffel 
symbols,
Ricci tensors and Ricci scalar, those used for the calculation. For an example, we have presented the $R_{++}$ 
equation of motion in detail in the appendix.

\section{The pp-wave background of ${AdS_4\times \C\cP^3}$} \label{pp wave background}
As a warm up exercise, in this section, we review few basic facts about the plane wave backgrounds obtained 
by taking Penrose limit of ${AdS_4\times \C\cP^3}$ \cite{Nishioka:2008gz,Grignani:2009ny,AliAkbari:2010rs}, which will
be helpful in fixing the notations etc. Depending on the number of space-like isometries, we have three different pp-wave
background arising from ${AdS_4\times \C\cP^3}$. Starting from the general form
of the metric, the three pp-wave backgrounds can be written down
by choosing appropriate values for different parameters.
The general form
of pp-wave geometry is given by \cite{Grignani:2009ny, AliAkbari:2010rs}, %
\begin{equation}
 \begin{split} %
 \label{ppg}
  ds^2&=-4dz^+dz^-+\sum_{{i}=1}^4\Big((du^{{i}})^2-(u^{{i}})^2(dz^+)^2\Big)
 +\sum_{j=1}^2\bigg[(dz^j)^2+(dy^j)^2  \cr %
 &+(\xi^2_j-\frac{1}{4})\left((z^j)^2+(y^j)^2\right)(dz^+)^2+2\bigg((\xi_j-2C_j)z^jdy^j-(\xi_j+2C_j)y^jdz^j\bigg)dz^+\bigg]
\end{split}
\end{equation}%
and by  choosing the $C_j$ and $\xi_j$ parameters appropriately, one arrives at the pp-wave background with 
no flat direction, one flat direction and two flat directions. In addition, there are two- and four-form
RR fields.\\
It is easy to see that if we choose the $C_j$ and $\xi_j$ parameters to be zero, i.e.
\begin{equation}
 \xi_j=C_j=0
 \label{pp0}
\end{equation}
the metric in eq(\ref{ppg}) will be reduced to the following form,
\begin{equation}\label{no flat}
ds^2=-4dz^+dz^--\left(\sum_{{i}=1}^4(u^i)^2+\frac{1}{4}\sum_{{j}=1}^{2}\left((z^{j})^{2}+(y^{j})^{2}\right)\right)
(dz^{+})^{2}+\sum_{{i}=1}^4 (du^i)^{2}+\sum_{{j}=1}^2 (dz^j)^{2}+\sum_{{j}=1}^2 (dy^j)^{2}
\end{equation}
This describes an  ${AdS_4\times \C\cP^3}$ pp-wave background with no explicit space-like isometry i.e. 
with no flat direction\cite{Nishioka:2008gz}.\\
Again, for the following choices of the  $C_j$ and $\xi_j$ parameters,
\begin{equation}
 \xi_1=\frac{1}{2},\xi_2=b+\frac{1}{2},C_1=\frac{1}{4},C_2=0,
 \end{equation}
 where $b$ is an arbitrary parameter, one gets the metric\cite{Grignani:2009ny} :
 \begin{equation}
 \label{one flat}
\begin{split}
ds^2=&-4dz^+dz^--\left(\sum_{i=1}^{4}(u^{i})^{2}-b(b+1)((z^{2})^{2}+(y^{2})^{2})\right)(dz^+)^2+(2b+1)
\left[z^2dy^2-y^2dz^2\right]dz^+
\\&-2y^1dz^1dz^++\sum_{i=1}^{4}(du^{i})^{2}+\sum_{j=1}^{2}(dz^{j})^{2}+\sum_{j=1}^{2}(dy^{j})^{2}
\end{split}
\end{equation}
From eq(\ref{one flat}), we find that $z^1$ is an explicit isometry of the background and hence this pp-wave 
background is referred as one flat direction pp-wave background.

If one chooses the $C_j$ and $\xi_j$ parameters to have the following values,
 \begin{equation}
  \xi_j=-\frac{1}{2},C_j=\frac{1}{4}
 \end{equation}
then the metric given in eq(\ref{ppg}) becomes
\begin{equation}\label{two flat}
\begin{split}
ds^2=&-4dz^+dz^--\sum_{i=1}^{4}(u^{i})^{2}(dz^+)^2+\sum_{i=1}^{4}(du^{i})^{2}+
\sum_{j=1}^{2}(dz^{j})^{2}+\sum_{j=1}^{2}(dy^{j})^{2}-2(z^{1}dy^1+z^2dy^2)dz^+
\end{split}
\end{equation}
The metric in eq(\ref{two flat}) describes exactly a pp-wave background with two flat directions\cite{Grignani:2009ny},
namely $y^1$ 
and $y^2$.

In the present work, we shall be interested in the pp-wave background with zero flat directions.
 By considering the following coordinate transformation,
\begin{eqnarray}
\nonumber
x^+=z^+\ \ \ \ \ \ \ \ \ \ x^-=-2z^-\ \ \ \ \ \ \ \ \ \ \ \  x^1=u^1\ \ \ \ \ \ \ \ \ \ 
x^2=u^2\ \ \ \ \ \ \ \ \ \ \  x^3=u^3\\
x^4=u^4\ \ \ \ \ \ \ \ \ \ \ \ \ 
x^5=z^1\ \ \ \ \ \ \ \ \ \ \ \ \ \ \  x^6=y^1\ \ \ \ \ \ \ \ \ \ \ x^7=z^2\ \ \ \ \ \ \ \ \ \ \ x^8=y^2
\end{eqnarray}

we can rewrite the ${AdS_4\times \C\cP^3}$ pp-wave background metric with zero 
flat directions given in eq(\ref{no flat}) as:

\begin{equation} 
 ds^2=2dx^+dx^--\left (\sum_{a=1}^{4} (x^a)^2+\frac{1}{4}\
\sum_{p=5}^{8}(x^p)^2 \right)(dx^+)^2+
\sum_{a=1}^{4}(dx^a)^2+\sum_{p=5}^{8}(dx^p)^2
\end{equation}

with the two- and four-form
RR fields $F_{+4}=\frac{k}{2\tilde{R}}$ and $F_{+123}=\frac{3}{2}\frac{k}{\tilde{R}}$, where $k$
is the level of the dual Chern-Simons gauge theory and $\tilde{R}$ is the radius of  ${AdS_4}$ and ${ \C\cP^3}$
as defined in \cite{Nishioka:2008gz}.
\footnote{see ref.\cite{Nishioka:2008gz} for details.}The dilaton is
expressed as $e^{\phi}=\frac{2\tilde{R}}{k}$.
Defining  $\mu = \frac{k}{2\tilde{R}}$, we can rewrite the values of
RR-fluxes as $F_{+4}=\mu$, $F_{+123}=3\mu$ and the dilaton as $ e^{\phi}= \mu^{-1} $.

For the subsequent analysis of the present paper, we shall be interested in the following background.
\begin{equation} \nonumber
 ds^2=2dx^+dx^--\left (\sum_{a=1}^{4} x_a^2+\frac{1}{4}\
\sum_{p=5}^{8}x_p^2 \right)(dx^+)^2+
\sum_{a=1}^{4}(dx^a)^2+\sum_{p=5}^{8}(dx^p)^2
\end{equation}
with \begin{equation} F_{+4}=\mu,\ \ \
F_{+123}=3\mu. \label{ppw}
\end{equation}
The light-cone superstring action on the above pp-wave background (\ref{ppw}) is
studied in \cite{Hyun:2002wu,Sugiyama:2002tf} and found to be preserving 24 supercharges.

\section{Supergravity Solution} \label{supergravity solution}
 In this section, we present the classical solution corresponding to 
 F-string  on the above pp-wave background (\ref{ppw}). We start by writing down
 the ansatz for the metric, the dilaton, the NS-NS  B-field and RR-field strengths of such a configuration,
\begin{eqnarray}
\nonumber
&& ds^2=f^{-1}\left[2dx^+dx^--\left( \sum_{a=1}^{4} (x^a)^2+\frac{1}{4}\
\sum_{p=5}^{8}(x^p)^2\right) (dx^+)^2\right]+
\sum_{a=1}^{4}(dx^a)^2+\sum_{p=5}^{8}(dx^p)^2 \\ \nonumber
&&e^{2\phi}=\mu^{-2}f^{-1},\\ \nonumber \ \ \ \
&&(F_2)_{+4}=\mu,\ \ \ \ \ \ \ \ \ \  (F_4)_{+123}=3\mu,  \\ \nonumber
&& H_{+-a}=\partial_af^{-1},\ \ \ \forall a=1,...4 \\
&&H_{+-p}=\partial _pf^{-1},\ \ \ \forall p= 5,...8 
\label{f-string solution}
\end{eqnarray}
where $f$ is a harmonic function in the
8-dimensional transverse space. We have explicitly verified that the above solution 
 satisfies  the type IIA field equations listed in \cite{Alishahiha:2000qf}. In particular, we found that
 the $R_{++}$ equation of motion will be satisfied if $ f$ satisfies the following condition,
 \footnote{see Appendix for details}
\begin{eqnarray}
 \sum_{a=1}^{4}\frac{\partial^2\textit{f}}{\partial x^{a2}}+\sum_{p=5}^{8}\frac{\partial^2\textit{f}}{\partial x^{p2}}=0
 \label{harmonic}
\end{eqnarray}

\section{Supersymmetry analysis} \label{supersymmtry analysis}
 In this section, we analyze the supersymmetry of the solution presented above in Section-{\ref{supergravity solution}}. 
 We start by  writing down
 the supersymmetry variations of dilatino and gravitino
fields of type IIA supergravity in ten dimensions, in string frame, \cite{Hassan:1999bv}  :

\begin{eqnarray} 
\label{dilatino}
\delta \lambda_{\pm} &=& {1\over2}(\Gamma^{\mu}\partial_{\mu}\phi \mp
{1\over 12} \Gamma^{\mu \nu \rho}H_{\mu \nu \rho})\epsilon_{\pm} + \\ \nonumber
&& {1\over
8}e^{\phi}(5 F^{0}\pm {3 \over 2!}\Gamma^{\mu \nu}F^{(2)}_{\mu \nu} + {1\over 4!} \Gamma^{\mu \nu
\rho\alpha}F^{(4)}_{\mu \nu \rho\alpha})\epsilon_{\mp},
\end{eqnarray}
\begin{eqnarray}
\delta {\Psi^{\pm}_{\mu}} &=& \Big[\partial_{\mu} + {1\over 4}(w_{\mu
\hat {\nu} \hat {\rho}} \mp {1\over 2} H_{\mu \hat{\nu}
\hat{\rho}})\Gamma^{\hat{\nu}\hat{\rho}}\Big]\epsilon_{\pm} \cr
& \cr
&+& {1\over 8}e^{\phi}\Big[F^{0}\pm {1\over 2!}  \Gamma^{\mu \nu }F^{(2)}_{\mu \nu} +{1\over 4!}
\Gamma^{\mu \nu \rho \alpha }F^{(4)}_{\mu \nu \rho \alpha
  }\Big]\Gamma_{\mu}\epsilon_{\mp},
\label{gravitino}
\end{eqnarray}
where we have used $(\mu, \nu ,\rho)$ to describe the ten
dimensional space-time indices, and hat's represent the corresponding
tangent space indices.

\subsection{ Background Supersymmetry} \label{background supersymmetry}
Before analyzing the supersymmetry of the F-string solution in the plane wave background of 
${AdS_4\times \C\cP^3}$ space-time, let's first discuss the supersymmetry of the background itself. 
The dilatino (\ref{dilatino}) and gravitino (\ref{gravitino}) variations impose non-trivial conditions 
on the spinor $\epsilon_{\pm}$.

Using the indices: $(+,-,a,p)$ to denote the ten dimensional coordinates with $a= 1,...,4$ and $ p = 5,...,8$,
we get the following condition from the dilatino equation for the ${AdS_4\times \C\cP^3}$ pp-wave background
given in eq(\ref{ppw})\footnote{Note that $F^{0}=0$ for the pp-wave background eq(\ref{ppw})under consideration.} 
(hats denoting the corresponding tangent space coordinates):
\begin{equation}
 \Big[ \Gamma^{\hat {+}\hat 4} \mp\Gamma^{\hat{+}\hat{1}\hat {2} 
\hat{3}}\Big] \epsilon_{\pm}=0
\label{bcgdilatino}
\end{equation}
Gravitino variation gives the following conditions on the spinors:
\footnote{We are in the frame where  $(\Gamma^{\hat{+}})^2= (\Gamma^{\hat{-}})^2=0 $, 
$[\Gamma^{\hat{+}},\Gamma^{\hat{-}}]_-= 2$. We have
made use of the identities: $\Gamma^{\hat{+}}\Gamma^{\hat{+}\hat{-}}
= - \Gamma^{\hat{+}}$, $\Gamma^{\hat{+}\hat{-}}\Gamma^{\hat{-}}
= - \Gamma^{\hat{-}}$ etc..}
\begin{eqnarray}
\label{bcggrav+}
&&\delta\psi_{+}^{\pm}=\partial_{+}\epsilon_{\pm}-\frac{1}{2}[x_{\hat{a}}\Gamma^{\hat{+}\hat{a}}+
\frac{1}{4}x_{\hat{p}}\Gamma^{\hat{+}\hat{p}}]\epsilon_{\pm}+
\frac{1}{8}[\pm\Gamma^{\hat{+}\hat{4}}+3\Gamma^{\hat{+}\hat{1}\hat{2}\hat{3}}]\Gamma^{\hat{-}}\epsilon_{\mp} =0\\
 \label{bcggrav-}
&& \delta\psi_{-}^{\pm}=\partial_{-}\epsilon_{\pm}=0 \\
\label{gravi}
&&\delta\psi_{a}^{\pm}=\partial_{a}\epsilon_{\pm}+
\frac{1}{8}[\pm\Gamma^{\hat{+}\hat{4}}+3\Gamma^{\hat{+}\hat{1}\hat{2}\hat{3}}]
\delta_{a\hat{a}}\Gamma^{\hat{a}}\epsilon_{\mp}=0.\ \ \ \forall a= 1,..4 \\
\label{grava}
&&\delta\psi_{p}^{\pm}=\partial_{p}\epsilon_{\pm}+
\frac{1}{8}[\pm\Gamma^{\hat{+}\hat{4}}+3\Gamma^{\hat{+}\hat{1}\hat{2}\hat{3}}]
\delta_{p\hat{p}}\Gamma^{\hat{p}}\epsilon_{\mp}=0.\ \ \ \forall p= 5,..8
\end{eqnarray}

Now there are two kinds of solutions of the above equations eq(\ref{bcgdilatino})- eq(\ref{grava}). 
One corresponds to considering $\Gamma^{\hat{+}}\epsilon_{\pm} = 0$, those are called `standard' or `normal'
Killing spinors \cite{Cvetic:2002si,SING,Alishahiha:2002rw}. This condition keeps 16 spinors out of the set of total 32.
Rest of the killing spinors for which  $\Gamma^{\hat{+}}\epsilon_{\pm} \neq 0$, are usually known as `supernumerary'
killing spinors.

Let's first consider the `normal' killing spinors with the condition
$\Gamma^{\hat{+}}\epsilon_{\pm} = 0$. For these spinors except eq\eqref{bcggrav+} all other equations can be
trivially satisfied. The eq(\ref{bcggrav+})- eq(\ref{grava}) are replaced by:
\begin{eqnarray}
\label{eq:gravitino plus pm}
&& \partial_{+}\epsilon_{\pm} - 
\frac{1}{4}[\pm\Gamma^{\hat{4}}+3\Gamma^{\hat{1}\hat{2}\hat{3}}]\epsilon_{\mp} = 0 \\
\label{gravitino minus,i}
&&\partial_{-}\epsilon_{\pm}=0, \>\>\>\>  \>\>\>\> \partial_{a}\epsilon_{\pm} = 0, 
\>\>\>\> \>\>\>\>\>  \partial_{p}\epsilon_{\pm} = 0
\end{eqnarray}
Eq\eqref{gravitino minus,i} implies that the spinors $\epsilon_{\pm}$ are independent of $x^-$, $x^a$ and $x^p$.
Eq\eqref{eq:gravitino plus pm} can now be solved easily after imposing either of the following two projections:
\begin{eqnarray}
 \Gamma^{\hat{1}\hat{2}\hat{3}\hat{4}}\epsilon_{\pm}= \pm\epsilon_{\pm}\label{first projection}\\
 \Gamma^{\hat{1}\hat{2}\hat{3}\hat{4}}\epsilon_{\pm}= \mp\epsilon_{\pm}\label{second projection}
 \end{eqnarray}
Explicitly, considering the projection eq\eqref{first projection}, eq\eqref{eq:gravitino plus pm} reduces to:
\begin{equation}\label{gravitino reduced from first projection}
\partial_{+}\epsilon_{\pm}-\Gamma^{\hat{1}\hat{2}\hat{3}}\epsilon_{\mp} = 0
\end{equation}
The Killing spinor equations are then solved by spinors: $\eta_{\pm}= \epsilon_{+} \pm\epsilon_{-}=
exp[\pm \Gamma^{\hat{1}\hat{2}\hat{3}}x^{+}]\eta_{0}$, with $\eta_{0}$ being a constant spinor. Similarly one can easily 
check that there are another set of spinors coming from the projection eq\eqref{second projection}.

Next we consider the `supernumerary' killing spinors with the condition
$\Gamma^{\hat{+}}\epsilon_{\pm} \neq 0$. These spinors will be constrained by the condition:
\begin{equation}\label{supernumerary condition}
[\pm\Gamma^{\hat{4}}+\Gamma^{\hat{1}\hat{2}\hat{3}}]\epsilon_{\mp}=0
\end{equation}
We now use  eq\eqref{supernumerary condition} to simplify the spinor equations eq\eqref{bcggrav+} -
eq\eqref{grava} further. The dilatino variation condition eq(\ref{bcgdilatino}) is now trivially satisfied 
and eqs(\ref{bcggrav+})- (\ref{grava}), following from $\delta {\Psi^{\pm}}_{\mu}=0$ can be written as:
\begin{eqnarray}
\label{reduced psi plus pm}
&&\partial_{+}\epsilon_{\pm}-\frac{1}{2}[x_{\hat{a}}\Gamma^{\hat{+}\hat{a}}+
\frac{1}{4}x_{\hat{p}}\Gamma^{\hat{+}\hat{p}}]\epsilon_{\pm}+
\frac{1}{4}\Gamma^{\hat{+}\hat{1}\hat{2}\hat{3}}\Gamma^{\hat{-}}\epsilon_{\mp} =0\\
 \label{reduced psi minus pm}
&&\partial_{-}\epsilon_{\pm}=0 \\
\label{psi a pm}
&&\partial_{a}\epsilon_{\pm}+\frac{1}{2}\Gamma^{\hat{+}\hat{1}\hat{2}\hat{3}}\delta_{a\hat{a}}\Gamma^{\hat{a}}
\epsilon_{\mp} = 0,
\ \ \ \forall a= 1,..4 \\
\label{psi p pm}
&&\partial_{p}\epsilon_{\pm}+
\frac{1}{4}\Gamma^{\hat{+}\hat{1}\hat{2}\hat{3}}\delta_{p\hat{p}} \Gamma^{\hat{p}}\epsilon_{\mp} = 0, 
\ \ \ \forall p= 5,..8
\end{eqnarray}
The above equations eq\eqref{reduced psi plus pm} - eq\eqref{psi p pm} can be solved for spinors
$\eta_{\pm}= \epsilon_{+} \pm\epsilon_{-}$. Adding the upper and lower sign equations in  
eq\eqref{reduced psi plus pm} - eq\eqref{psi p pm} lead to the differential equations for $\eta_+$, 
while substracting the two give those for $\eta_-$. The gravitini variations in terms of  $\eta_+$  can be written as:
\begin{eqnarray}
\label{eta+ eqn +}
&& \partial_{+}\eta_{+}-\frac{1}{2}[x_{\hat{a}}\Gamma^{\hat{+}\hat{a}}+
\frac{1}{4}x_{\hat{p}}\Gamma^{\hat{+}\hat{p}}]\eta_{+}+
\frac{1}{4}\Gamma^{\hat{+}\hat{1}\hat{2}\hat{3}}\Gamma^{\hat{-}}\eta_{+}=0\\
\label{eta+ eqn a}
&&\partial_{a}\eta_{+}+\frac{1}{2}\Gamma^{\hat{+}\hat{1}\hat{2}\hat{3}}\delta_{a\hat{a}}\Gamma^{\hat{a}}\eta_{+}=0\\
\label{eta+ eqn p}
&&\partial_{p}\eta_{+}+\frac{1}{4}\Gamma^{\hat{+}\hat{1}\hat{2}\hat{3}}\delta_{p\hat{p}}\Gamma^{\hat{p}}\eta_{+}=0
\end{eqnarray}
The $x^-$ component of the Killing spinor equation eq\eqref{reduced psi minus pm} is found to be reduced
to $\partial_{-}\eta_{+} =0 $,
which implies that $\eta_+$ is independent of $x^-$. The $ x^a$ and $x^p$ components of the Killing spinor equation
can be written as 
$(\partial_{a}+\Omega_{a})\eta_{+} = 0 $ and $\left(\partial_{p}+\Omega_{p}\right)\eta_{+}=0$ respectively.  
Since $\Omega_{a}\Omega_{b}=0 = \Omega_{p}\Omega_{q}$
for any $a,b = 1\cdots 4 $ and any $p,q = 5\cdots 8 $  using the condition $(\Gamma^{\hat{+}})^2=0 $, the $x^a$ as
well as $x^p$
dependence of  $\eta_+$ 
is expected to be linear.
The solution of the above Killing spinor equations
eq\eqref{eta+ eqn +} - eq\eqref{eta+ eqn p}
is then found in a similar way as in \cite{Cvetic:2002si,Alishahiha:2002rw,meessen} and is given as:
\begin{equation}
 \eta_{+}=\left(1-\Omega_{a}x^{a} -\Omega_{p}x^p\right)
exp\left[- \frac{1}{4}\Gamma^{\hat{+}\hat{1}\hat{2}\hat{3}}\Gamma^{\hat{-}}x^{+}\right].\eta_0
\label{solution eta+}
\end{equation}
with $\eta_0$ being a constant spinor and $\Omega_{a}$'s, $\Omega_{p}$'s are given by,
\begin{equation}
  \Omega_{a}=\frac{1}{2}\Gamma^{\hat{+}}\Gamma^{\hat{1}\hat{2}\hat{3}}\delta_{a\hat{a}}\Gamma^{\hat{a}}, \ \ \
  \Omega_{p}=\frac{1}{4}\Gamma^{\hat{+}}\Gamma^{\hat{1}\hat{2}\hat{3}}\delta_{p\hat{p}}\Gamma^{\hat{p}}
  \label{omegas}
\end{equation}

One can easily write down the gravitini variations in terms of  $\eta_-$ and verify that the spinor equations are 
satisfied with spinors: $\eta_{-}=\left(1+\Omega_{a}x^{a}+\Omega_{p}x^p\right)
exp\left[ \frac{1}{4}\Gamma^{\hat{+}\hat{1}\hat{2}\hat{3}}\Gamma^{\hat{-}}x^{+}\right].\eta_0$.

The 16 standard Killing spinors  and  8 supernumerary ones give a total  of 24 Killing spinors for the 
type IIA pp-wave background presented in eq\eqref{ppw}.
Thus the total number of supersymmetries preserved by the background pp-wave will be $\frac{3}{4}$ of the maximal ones.
We thus confirm the results discussed in \cite{Hyun:2002wu,Sugiyama:2002tf}.

\subsection{F-string supersymmetry} \label{f-string supersymmtry}
In this section we will present the supersymmetry of the F-string solution described earlier in
section-\ref{supergravity solution}. We will see below that in spite of its simple form, this solution does not preserve
any space-time supersymmetry. Solving the supersymmetry variations of dilatino and gravitino fields given in 
eq\eqref{dilatino} - eq\eqref{gravitino} for the solution describing an F-string in eq\eqref{f-string solution}
\footnote{Here also we have $F^{0}=0$},
 we get 
several conditions. The 
vanishing of the dilatino variations give
\begin{eqnarray} 
\delta \lambda_{\pm} &=& {f_{,\hat a}\over4f}\Big[-\Gamma^{\hat a} \pm \Gamma^{\hat{+}\hat{-}\hat a}\Big] 
\epsilon_{\pm} +  {f_{,\hat p}\over4f}\Big[-\Gamma^{\hat p} \pm \Gamma^{\hat{+}\hat{-}\hat p}\Big] 
\epsilon_{\pm} + {3 \over 8 } \Big[\pm \Gamma^{\hat {+}\hat 4} +
\Gamma^{\hat{+}\hat{1}\hat {2} \hat{3}}\Big] \epsilon_{\mp} =0.
\label{fstringdilatino}
\end{eqnarray}
Eq\eqref{fstringdilatino} to be hold, we should have the following  conditions satisfied:
\begin{eqnarray}
\label{extracondition}
 && \Big[\Gamma^{\hat a} \mp \Gamma^{\hat{+}\hat{-}\hat a} \Big] \epsilon_{\pm} = 0, \ \ \ \ \ \ \ \ \ \
 \Big[\Gamma^{\hat p} \mp \Gamma^{\hat{+}\hat{-}\hat p} \Big] \epsilon_{\pm} = 0\\
 \label{usualdilatino}
 &&\Big[ \Gamma^{\hat {+}\hat 4} \mp\Gamma^{\hat{+}\hat{1}\hat {2} 
\hat{3}}\Big] \epsilon_{\pm}=0 
 \end{eqnarray}
 Gravitino variation gives the following conditions on the spinors:
 \begin{eqnarray}
\label{fstringgrav+}
&&\delta\psi_{+}^{\pm}=\partial_{+}\epsilon_{\pm}-\frac{1}{2f^{1/2} }[x_{\hat{a}}\Gamma^{\hat{+}\hat{a}}+
\frac{1}{4}x_{\hat{p}}\Gamma^{\hat{+}\hat{p}}]\epsilon_{\pm}+
\frac{1}{8 f^{1/2}}[\pm\Gamma^{\hat{+}\hat{4}}+3\Gamma^{\hat{+}\hat{1}\hat{2}\hat{3}}]\Gamma^{\hat{-}}\epsilon_{\mp} =0\\
 \label{fstringgrav-}
&& \delta\psi_{-}^{\pm}=\partial_{-}\epsilon_{\pm}=0 \\
\label{fstringgravi}
&&\delta\psi_{a}^{\pm}=\partial_{a}\epsilon_{\pm} + \frac{1}{4}\frac{f_{,a}}{f} \epsilon_{\pm} +
\frac{1}{8}[\pm\Gamma^{\hat{+}\hat{4}}+3\Gamma^{\hat{+}\hat{1}\hat{2}\hat{3}}]
\delta_{a\hat{a}}\Gamma^{\hat{a}}\epsilon_{\mp}=0.\ \ \ \forall a= 1,..4 \\
\label{fstringgrava}
&&\delta\psi_{p}^{\pm}=\partial_{p}\epsilon_{\pm} + \frac{1}{4}\frac{f_{,p}}{f} \epsilon_{\pm} +
    \frac{1}{8}[\pm\Gamma^{\hat{+}\hat{4}}+3\Gamma^{\hat{+}\hat{1}\hat{2}\hat{3}}]
    \delta_{p\hat{p}}\Gamma^{\hat{p}}\epsilon_{\mp}=0.\ \ \ \forall p= 5,..8  
\end{eqnarray}
In writing the above set of conditions, we have used eq\eqref{extracondition}.

First, looking for the ` normal' Killing spinors satisfying the condition $\Gamma^{\hat{+}}\epsilon_{\pm}=0$. 
In addition, we have to use either of the  two projections given in eq\eqref{first projection} and
eq\eqref{second projection}.
Proceeding with the projection condition eq\eqref{first projection}, the gravitini equations from eq\eqref{fstringgrav+} -
eq\eqref{fstringgrava} 
to be solved are:
\begin{eqnarray}
\label{fstringnormal+}
 &&\partial_{+}\epsilon_{\pm}- \frac{1}{f^{1/2} }\Gamma^{\hat{1}\hat{2}\hat{3}}\epsilon_{\mp} = 0,\\
 \label{fstringnormali}
 &&\partial_{a}\epsilon_{\pm} + \frac{1}{4}\frac{f_{,a}}{f} \epsilon_{\pm} = 0, \ \ \ \forall a= 1,..4 \\ 
 \label{fstringnormala}
 &&\partial_{p}\epsilon_{\pm}+\frac{1}{4}\frac{f_{,p}}{f}\epsilon_{\pm}=0, \ \ \ \forall p= 5,..8  
\end{eqnarray}

Adding the upper and lower sign equations in eq\eqref{fstringnormal+} - eq\eqref{fstringnormala}, we have 
the killing spinor equations for $\eta_{+}= \epsilon_{+} +\epsilon_{-}$ and 
substracting them we get the corresponding ones for $\eta_{-}= \epsilon_{+} - \epsilon_{-}$. Let us first consider
the gravitini variations for $\eta_+$:

\begin{eqnarray}
\label{normaleta+}
 &&\partial_{+}\eta_+ - \frac{1}{f^{1/2} }\Gamma^{\hat{1}\hat{2}\hat{3}}\eta_+ = 0,\\
 \label{normaletaa}
 &&\partial_{a}\eta_+ + \frac{1}{4}\frac{f_{,a}}{f} \eta_+ = 0 \\
 \label{normaletap}
  &&\partial_{p}\eta_{+}+\frac{1}{4}\frac{f_{,p}}{f}\eta_{+}=0
\end{eqnarray}
One can check that eq\eqref{normaleta+} - eq\eqref{normaletap} are non integrable ones due to the specific form of 
the function $f$. We have explicitly checked that  the equations for $\eta_{-}$ are also non integrable.

Next, we will analyze the `supernumerary' Killing spinors. The vanishing of the dilatino variation implies 
that these spinors will be constrained by eq\eqref{supernumerary condition} in addition to eq\eqref{extracondition}.
The gravitini variations eq\eqref{fstringgrav+} - eq\eqref{fstringgrava} can be simplified further using 
eq\eqref{supernumerary condition} and eq\eqref{extracondition}
and can be written as :

\begin{eqnarray}
\label{fstringsupernumerary+}
&&\partial_{+}\epsilon_{\pm}-\frac{1}{2f^{1/2} }[x_{\hat{a}}\Gamma^{\hat{+}\hat{a}}+
\frac{1}{4}x_{\hat{p}}\Gamma^{\hat{+}\hat{p}}]\epsilon_{\pm}+
\frac{1}{4 f^{1/2}}\Gamma^{\hat{+}\hat{1}\hat{2}\hat{3}}\Gamma^{\hat{-}}\epsilon_{\mp} =0\\
 \label{fstringsupernumerary-}
&& \partial_{-}\epsilon_{\pm}=0 \\
\label{fstringsupernumerary-a}
&&\partial_{a}\epsilon_{\pm}+  \frac{1}{4}\frac{f_{,a}}{f} \epsilon_{\pm} +
\frac{1}{2}\Gamma^{\hat{+}\hat{1}\hat{2}\hat{3}}\delta_{a\hat{a}}\Gamma^{\hat{a}}
\epsilon_{\mp} = 0,
\ \ \ \forall a= 1,..4 \\
\label{fstringsupernumerary-p}
&&\partial_{p}\epsilon_{\pm}+  \frac{1}{4}\frac{f_{,p}}{f} \epsilon_{\pm} +
\frac{1}{4}\Gamma^{\hat{+}\hat{1}\hat{2}\hat{3}}\delta_{p\hat{p}} \Gamma^{\hat{p}}\epsilon_{\mp} = 0, 
\ \ \ \forall p= 5,..8
\end{eqnarray}
Now,one can easily write down the equations in the basis  $\eta_{\pm}= \epsilon_{+} \pm\epsilon_{-}$ and check that 
 these are non integrable ones due to specific dependence of $f$. This shows that the F-string solution in equation 
 \eqref{f-string solution} does not preserve any of the space-time supersymmetries.

\section{Summary and Discussion} \label{conclusion}

In this paper we have found the supergravity solution of fundamental string in the pp-wave background arising from 
the Penrose limit of ${AdS_4\times \C\cP^3}$ with zero space like isometries presented in \cite{Nishioka:2008gz}.
We discussed the supersymmetry of the pp-wave background and also the  F- string solution by solving the type-II 
Killing spinor equations explicitly. We confirmed that the background preserves 24 Killing spinors  out of which
16 are of 'standard' type and rest 8 are 'supernumerary' in nature. It is shown that the classical F-string solution 
on this background does not preserve any space-time supersymmetry. The corresponding
worldsheet analysis may reveal the reason for the solution destroying all space-time supersymmetry.
There is a possibility that while supersymmetry is preserved on the worldsheet
it does not admit a local space-time realization.\footnote{ We 
thank the unknown reviewer for the useful discussions on this issue.}
One could possibly try to study the stability of the background. It is interesting to investigate the
F-string solution in the pp-wave backgrounds with one flat direction and two flat directions. It might also be
possible to use series of T- and S- duality to generate other solutions for Dp-branes in this background.
We hope
to return to these questions in our future exercise. The solution obtained in this paper seems very 
interesting in the line of supergravity theories. However, the physical significance of the solution remain
elusive from us as of now.

\section*{Acknowledgment}
This work is dedicated to late Prof. Alok Kumar.
It is a great pleasure to thank Mohsen Alishahiha for valuable
discussions, comments and for
reading the manuscript carefully. We would like to thank Harvendra Singh, Sayantani Bhattacharyya and Sachin Jain 
for useful discussions. 
 B.P. acknowledges IIT(ISM), Dhanbad for the Grant (FRS (53)/2013-2014/APH).

\section{Appendix} \label{appendix}
 
In this appendix, we summarize the nonvanishing christoffel symbols, Ricci tensors and Ricci scalar
obtaind from the eq\eqref{f-string solution}:
 \begin{eqnarray}
 \Gamma^+_{+a}=-\frac{1}{2f}\frac{\partial f}{\partial x^a}, &
  \Gamma^+_{+p}=-\frac{1}{2f}\frac{\partial f}{\partial x^p}, &\Gamma^-_{+a}=-x^a, \\
 \Gamma^a_{++}=\frac{x^a}{f}-\frac{1}{2f^2}\frac{\partial f}{\partial x^a}
  \left[\sum_{a=1}^{4}\left(x^a\right)^2+\frac{1}{4}\sum_{p=5}^{8}\left(x^p\right)^2\right], 
  & \Gamma^-_{-a}=-\frac{1}{2f} \frac{\partial f}{\partial x^a}, 
  &\Gamma^-_{-p}=-{\frac{1}{2f}\frac{\partial f}{\partial x^p}}\\
  \Gamma^p_{++}=\frac{x^p}{4f}-\frac{1}{2f^2}\frac{\partial f}{\partial x^p}\left[\sum_{a=1}^{4}\left(x^a\right)^2+
\frac{1}{4}\sum_{p=5}^{8}\left(x^p\right)^2\right]
& \Gamma^a_{+-}=\frac{1}{2}\frac{1}{f^2}\frac{\partial f}{\partial x^a},
&\Gamma^p_{+-}=\frac{1}{2}\frac{1}{f^2}\frac{\partial f}{\partial x^p}\\
\Gamma^-_{+p} = -\frac{x^p}{4} &&
\end{eqnarray}
And the Ricci tensors thus  obtained are given by:
  \begin{equation}
 \begin{split}
 R_{++} = \frac{1}{f^{3}}\left(\sum_{a=1}^{4}x^{a2}+\sum_{p=5}^{8}\frac{x^{p2}}{4}\right)
 \left[\sum_{a=1}^{4}\left(\frac{\partial f}{\partial x^{a}}\right)^2+\sum_{p=5}^{8}
          \left(\frac{\partial f}{\partial x^{p}}\right)^2\right]\\
 -\frac{1}{2f^{2}}\left(\sum_{a=1}^{4}x^{a2}+\sum_{p=5}^{8}\frac{x^{p2}}{4}\right)
\left[\sum_{a=1}^{4}\frac{\partial^2\textit{f}}{\partial x^{a2}}+\sum_{p=5}^{8}\frac{\partial^2\textit{f}}
{\partial x^{p2}}\right] \\
 -\frac{1}{f^{2}}\left[\sum_{a=1}^{4}x^{a}\left(\frac{\partial f}{\partial x^{a}}\right)+\sum_{p=5}^{8}\frac{x^p}{4}
 \left(\frac{\partial f}{\partial x^p}\right)\right]+\frac{5}{f} 
 \end{split}
 \label{lhs}
\end{equation}
\begin{equation}
 R_{+-}={\frac{1}{2f^2}\left[\sum_{a=1}^{4}\frac{\partial^2\textit{f}}{\partial x^{a2}}+\sum_{p=5}^{8}\frac{\partial^2\textit{f}}
{\partial x^{p2}}\right]-
 \frac{1}{f^3}\left[\sum_{a=1}^{4}\left(\frac{\partial f}{\partial x^{a}}\right)^2+\sum_{p=5}^{8}
          \left(\frac{\partial f}{\partial x^{p}}\right)^2\right]}
\end{equation}
\begin{equation}
 R_{ab}={\frac{1}{f}\left[\sum_{a,b=1}^{4}\frac{\partial}{\partial x^{b}}
 \left(\frac{\partial f}{\partial x^{a}}\right)\right]-\frac{3}{2f^2}
 \left[\sum_{a,b=1}^{4}\left(\frac{\partial f}{\partial x^a}\right)\left(\frac{\partial f}{\partial x^b}\right)\right]}
\end{equation}
\begin{equation}
 R_{pq}={\frac{1}{f}\left[\sum_{p,q=5}^{8}\frac{\partial }{\partial x^{q}}
 \left(\frac{\partial f}{\partial x^{p}}\right)\right]-\frac{3}{2f^2}
 \left[\sum_{p,q=5}^{8}\left(\frac{\partial f}{\partial x^p}\right)\left(\frac{\partial f}{\partial x^q}\right)\right]}
\end{equation}
\begin{equation}
R_{ap}=-\frac{3}{2f^2}\left[\sum_{a=1}^{4}\frac{\partial f}{\partial x^a}
\sum_{p=5}^{8}\frac{\partial f}{\partial x^p}\right]+
 \frac{1}{f}\left[\sum_{a=1}^{4}\sum_{p=5}^{8}\frac{\partial}{\partial x^p}
 \left(\frac{\partial f}{\partial x^a}\right)\right]
 \end{equation}
 The Ricci scalar is given as:\\
  \begin{equation}
  R=-\frac{7}{2f^2}\left[\sum_{a=1}^{4}\left(\frac{\partial f}{\partial x^a}\right)^2+\sum_{p=5}^{8}\left(\frac{\partial f}{\partial x^p}\right)^2\right]
+\frac{2}{f}\left[\sum_{a=1}^{4}\left(\frac{\partial ^2f}{\partial x^{a2}}\right)+\sum_{p=5}^{8}\left(\frac{\partial ^2f}{\partial x^{p2}}\right)\right]
 \end{equation}
 
The background  with the ansatz for the  metric, the dilaton, the NS-NS  B-field and RR field strengths,
described in eq\eqref{f-string solution} satisfies the type IIA field equations listed in \cite{Alishahiha:2000qf}.
For an example let's consider the $R_{++}$ equation of motion explicitly. Solving eq(67) of  \cite{Alishahiha:2000qf}
for $i,j = +$ with the values for the dilaton, the NS-NS  B-field and RR field strengths as given 
in eq\eqref{f-string solution}, one gets:
 \begin{equation}
 \begin{split}
 R_{++} =& \frac{1}{f^{3}}\left(\sum_{a=1}^{4}x^{a2}+\sum_{p=5}^{8}\frac{x^{p2}}{4}\right)
 \left[\sum_{a=1}^{4}\left(\frac{\partial f}{\partial x^{a}}\right)^2+\sum_{p=5}^{8}\left(\frac{\partial f}
 {\partial x^{p}}\right)^2\right]\\&
  -\frac{1}{f^{2}}\left[\sum_{a=1}^{4}x^{a}\left(\frac{\partial f}{\partial x^{a}}\right)+\sum_{p=5}^{8}\frac{x^p}{4}
 \left(\frac{\partial f}{\partial x^p}\right)\right]+\frac{5}{f} 
  \end{split}
  \label{rhs}
  \end{equation}
  Clearly, eq(\ref{lhs}) is in exact match with eq(\ref{rhs}) when $ f$ satisfies eq(\ref{harmonic}).


\begin{thebibliography}{99}
\bibitem{penrose} R. Penrose, {\it ``Any space-time has a plane wave as a
  limit''},in Differential geometry and relativity, pp.271-275,
  Reidel, Dordrecht, (1976).
  
   \bibitem{Horowitz:1994rf} 
  G.~T.~Horowitz and A.~A.~Tseytlin,
  ``A New class of exact solutions in string theory,''
  Phys.\ Rev.\ D {\bf 51}, 2896 (1995)
  [hep-th/9409021].
  
  \bibitem{Gueven:2000ru} 
  R.~Gueven,
  ``Plane wave limits and T duality,''
  Phys.\ Lett.\ B {\bf 482}, 255 (2000)
  [hep-th/0005061].
 \bibitem{Metsaev:1999gz} 
 R.R.Metsaev, {\it ``Light cone gauge formulation 
of IIB supergravity in $AdS_5 \times S^5$ background and 
AdS/CFT correspondence",} Phys.Lett. {\bf B468} (1999) 65, hep-th/9908114.
 \bibitem{Metsaev:2001bj}
R.R.Metsaev, {\it ``Type IIB Green-Schwarz
superstring in plane wave Ramond-Ramond background",}Nucl. Phys. {\bf
B625}, (2002) 70, hep-th/0112044.
\bibitem{Metsaev:2002re}
R.R. Metsaev, A.A. Tseytlin, {\it ``Exactly solvable model of
superstring in Ramond-Ramond plane wave background'',}
Phys.Rev. {\bf D65} (2002) 126004, hep-th/0202109.
 \bibitem{Blau:2001ne} 
  M.Blau, J.Figuero-O'Farrill, C. Hull and 
G. Papadopoulos,{\it `` A new maximally supersymmetric background of
IIB superstring theory,''} JHEP {\bf 0201}, 047 (2000), hep-th/0110242
\bibitem{Blau:2002dy}
M.Blau, J.Figuero-O'Farrill and G. Papadopoulos,{\it `` 
Penrose limits and maximal supersymmetry,''} Class. Quant. Grav.
{\bf 19, L87}(2000), hep-th/0201081.
\bibitem{Blau:2002mw}
M.Blau, J.Figuero-O'Farrill and G. Papadopoulos,{\it
``Penrose limits, supergravity and brane dynamics,''}
hep-th/0202111.
  
 
\bibitem{Berenstein:2002jq} 
  D.~E.~Berenstein, J.~M.~Maldacena and H.~S.~Nastase,
  JHEP {\bf 0204}, 013 (2002)
  doi:10.1088/1126-6708/2002/04/013
  [hep-th/0202021].
 
 
 
 \bibitem{mukhi} N.~Itzhaki, I.~R.~Klebanov and S.~Mukhi,
{\it ``PP wave limit and enhanced supersymmetry in gauge theories,''}
JHEP {\bf 0203}, 048 (2002), hep-th/0202153.
 
 \bibitem{GO} J.~Gomis and H.~Ooguri,{\it ``Penrose limit of N = 1 
gauge theories",} hep-th/0202157.

\bibitem{tseytlin} J.G. Russo, A.A. Tseytlin, {\it ``On solvable 
models of type IIB superstring in NS-NS and R-R plane wave
backgrounds,"} JHEP {\bf 0204} (2002) 021, hep-th/0202179.

\bibitem{sonn} L.~A. Pando~Zayas and J.~Sonnenschein,
{\it ``On Penrose limits and gauge theories,''} hep-th/0202186.
 
 \bibitem{mohsen} M. Alishahiha and M. M. Sheikh-Jabbari,
{\it ``The PP-wave limits of orbifolded $AdS_5 \times S^5$,''}
hep-th/0203018.

\bibitem{chu} C. S. Chu, P. M. Ho, {\it ``Noncommutative D-brane and 
Open String in pp-wave Background with B-field,''}
Nucl.Phys. {\bf B636} (2002) 141, hep-th/0203186.

\bibitem{lee} P.~Lee and J.~W.~Park,
{\it ``Open strings in PP-wave background from defect conformal field
theory'',} hep-th/0203257.

\bibitem{Bala} V.~Balasubramanian, M.~X.~Huang, T.~S.~Levi and A.~Naqvi,
{\it ``Open strings from N = 4 super Yang-Mills,''} hep-th/0204196.

\bibitem{Taka} H.~Takayanagi and T.~Takayanagi,
{\it ``Open strings in exactly solvable model of curved space-time and
PP-wave limit,''}hep-th/0204234.
 
 \bibitem{meessen} P. Meessen, {\it ``A Small Note on PP-Wave Vacua 
in 6 and 5 Dimensions,"} Phys.Rev.{\bf D65}(2002) 087501,
hep-th/0111031.

\bibitem{pope}  M. Cvetic, H. Lu, C.N. Pope,{\it ``Penrose Limits, 
PP-Waves and Deformed M2-branes,"} hep-th/0203082.

\bibitem{hull}  J. P. Gauntlett, C. M. Hull, {\it ``pp-waves in
    11-dimensions with extra supersymmetry",}JHEP {\bf 0206}, (2002)
  013, hep-th/0203255.
\bibitem{Biswas:2002yz}
  A.~Biswas, A.~Kumar and K.~L.~Panigrahi,
  ``P - p-prime branes in PP wave background,''
  Phys.\ Rev.\ D {\bf 66} (2002) 126002
  doi:10.1103/PhysRevD.66.126002
  [hep-th/0208042].
 \bibitem{Kumar:2002ps}  
   A.~Kumar, R.~R.~Nayak, S.~Siwach and ,
  ``D-brane solutions in p p wave background,''
  Phys.\ Lett.\ B {\bf 541}, 183 (2002)
  [hep-th/0204025].
  \bibitem{Alishahiha:2002zu}
    M.~Alishahiha and A.~Kumar,
  ``PP waves from nonlocal theories,''
  JHEP {\bf 0209}, 031 (2002)
   [hep-th/0207257].
   \bibitem{Hassan:2003ec}
     S.~F.~Hassan, R.~R.~Nayak and K.~L.~Panigrahi,
  ``D branes in the NS5 near horizon pp wave background,''
  hep-th/0312224.
\bibitem{Alishahiha:2002rw}
  M.~Alishahiha and A.~Kumar,
  ``D-brane solutions from new isometries of pp waves,''
  Phys.\ Lett.\ B {\bf 542} (2002) 130
    [hep-th/0205134]
 
 \bibitem{Nishioka:2008gz} 
  T.~Nishioka and T.~Takayanagi,
  ``On Type IIA Penrose Limit and N=6 Chern-Simons Theories,''
  JHEP {\bf 0808}, 001 (2008)
  [arXiv:0806.3391 [hep-th]].
  \bibitem{Maldacena:1997re}
  J.~M.~Maldacena,
  ``The large N limit of superconformal field theories and supergravity,''
  Adv.\ Theor.\ Math.\ Phys.\  {\bf 2}, 231 (1998)
  [Int.\ J.\ Theor.\ Phys.\  {\bf 38}, 1113 (1999)]
  [arXiv:hep-th/9711200].
\bibitem{Witten:1998qj}
  E.~Witten,
  ``Anti-de Sitter space and holography,''
  Adv.\ Theor.\ Math.\ Phys.\  {\bf 2}, 253 (1998)
  [arXiv:hep-th/9802150]
\bibitem{Gubser:1998bc}
  S.~S.~Gubser, I.~R.~Klebanov and A.~M.~Polyakov,
  ``Gauge theory correlators from non-critical string theory,''
  Phys.\ Lett.\  B {\bf 428}, 105 (1998)
  [arXiv:hep-th/9802109].
\bibitem{Schwarz:2004yj}
  J.~H.~Schwarz,
  ``Superconformal Chern-Simons theories,''
  JHEP {\bf 0411}, 078 (2004)
  [arXiv:hep-th/0411077].
  \bibitem{Bagger:2006sk}
 J.~Bagger and N.~Lambert,
  ``Modeling multiple M2's,''
  Phys.\ Rev.\  D {\bf 75}, 045020 (2007)
  [arXiv:hep-th/0611108].
  \bibitem{Bagger:2007jr}
J.~Bagger and N.~Lambert,
  ``Gauge Symmetry and Supersymmetry of Multiple M2-Branes,''
  Phys.\ Rev.\  D {\bf 77}, 065008 (2008)
  [arXiv:0711.0955 [hep-th]].
  \bibitem{Gustavsson:2007vu} 
  A.~Gustavsson,
  ``Algebraic structures on parallel M2-branes,''
  [arXiv:0709.1260 [hep-th]].
  \bibitem{Gomis:2008be}
  J.~Gomis, D.~Rodriguez-Gomez, M.~Van Raamsdonk and H.~Verlinde,
  ``Supersymmetric Yang-Mills Theory From Lorentzian Three-Algebras,''
  JHEP {\bf 0808}, 094 (2008)
  [arXiv:0806.0738 [hep-th]].
\bibitem{Bagger:2007vi}
  J.~Bagger and N.~Lambert,
  ``Comments On Multiple M2-branes,''
  JHEP {\bf 0802}, 105 (2008)
  [arXiv:0712.3738 [hep-th]].
 \bibitem{Gaiotto:2007xh} 
  D.~Gaiotto and X.~Yin,
  ``Genus Two Partition Functions of Extremal Conformal Field Theories,''
  JHEP {\bf 0708}, 029 (2007)
  [arXiv:0707.3437 [hep-th]].
\bibitem{Gaiotto:2008sd}
  D.~Gaiotto and E.~Witten,
  ``Janus Configurations, Chern-Simons Couplings, And The Theta-Angle in N=4
  [arXiv:0804.2907 [hep-th]].
\bibitem{Bandres:2008vf}
  M.~A.~Bandres, A.~E.~Lipstein and J.~H.~Schwarz,
  ``N = 8 Superconformal Chern--Simons Theories,''
  JHEP {\bf 0805}, 025 (2008)
  [arXiv:0803.3242 [hep-th]].
\bibitem{VanRaamsdonk:2008ft}
  M.~Van Raamsdonk,
  ``Comments on the Bagger-Lambert theory and multiple M2-branes,''
  JHEP {\bf 0805}, 105 (2008)
  [arXiv:0803.3803 [hep-th]].
\bibitem{Lambert:2008et}
  N.~Lambert and D.~Tong,
  ``Membranes on an Orbifold,''
  Phys.\ Rev.\ Lett.\  {\bf 101}, 041602 (2008)
  [arXiv:0804.1114 [hep-th]].
\bibitem{Benvenuti:2008bt}
  S.~Benvenuti, D.~Rodriguez-Gomez, E.~Tonni and H.~Verlinde,
  ``N=8 superconformal gauge theories and M2 branes,''
  [arXiv:0805.1087 [hep-th]].
\bibitem{Krishnan:2008zm}
  C.~Krishnan and C.~Maccaferri,
  ``Membranes on Calibrations,''
  JHEP {\bf 0807}, 005 (2008)
  [arXiv:0805.3125 [hep-th]].
\bibitem{Bandres:2008kj}
  M.~A.~Bandres, A.~E.~Lipstein and J.~H.~Schwarz,
  ``Ghost-Free Superconformal Action for Multiple M2-Branes,''
  JHEP {\bf 0807}, 117 (2008)
  [arXiv:0806.0054 [hep-th]].
\bibitem{Bagger:2008se}
  J.~Bagger and N.~Lambert,
  ``Three-Algebras and N=6 Chern-Simons Gauge Theories,''
  [arXiv:0807.0163 [hep-th]].

  
\bibitem{Aharony:2008ug} 
  O.~Aharony, O.~Bergman, D.~L.~Jafferis and J.~Maldacena,
  ``N=6 superconformal Chern-Simons-matter theories, M2-branes and their gravity duals,''
  JHEP {\bf 0810}, 091 (2008)
  [arXiv:0806.1218 [hep-th]].
  
  \bibitem{Arutyunov:2008if}
  G.~Arutyunov and S.~Frolov,
  ``Superstrings on $AdS_4 x CP^3$ as a Coset Sigma-model,''
  JHEP {\bf 0809}, 129 (2008)
  [arXiv:0806.4940 [hep-th]].
\bibitem{Stefanski:2008ik}
  B.~.~j.~Stefanski,
  ``Green-Schwarz action for Type IIA strings on $AdS_4\times CP^3$,''
  Nucl.\ Phys.\  B {\bf 808}, 80 (2009)
  [arXiv:0806.4948 [hep-th]].
\bibitem{Bonelli:2008us}
  G.~Bonelli, P.~A.~Grassi and H.~Safaai,
  ``Exploring Pure Spinor String Theory on $AdS_4\times \mathbb{CP}^3$,''
  JHEP {\bf 0810}, 085 (2008)
  [arXiv:0808.1051 [hep-th]].
\bibitem{D'Auria:2008cw}
  R.~D'Auria, P.~Fre, P.~A.~Grassi and M.~Trigiante,
  ``Superstrings on $AdS_4 x CP^3$ from Supergravity,''
  Phys.\ Rev.\  D {\bf 79}, 086001 (2009)
  [arXiv:0808.1282 [hep-th]].
\bibitem{Sorokin:1985ap}
  D.~P.~Sorokin, V.~I.~Tkach and D.~V.~Volkov,
  ``On The Relationship Between Compactified Vacua Of D = 11 And D = 10
  Supergravities,''
  Phys.\ Lett.\  B {\bf 161}, 301 (1985).
\bibitem{Gomis:2008jt}
  J.~Gomis, D.~Sorokin and L.~Wulff,
  ``The complete $AdS(4) x CP(3)$ superspace for the type IIA superstring and
  D-branes,''
  JHEP {\bf 0903}, 015 (2009)
  [arXiv:0811.1566 [hep-th]].
\bibitem{Grassi:2009yj}
  P.~A.~Grassi, D.~Sorokin and L.~Wulff,
  ``Simplifying superstring and D-brane actions in $AdS(4) x CP(3)$
  superbackground,''
  JHEP {\bf 0908}, 060 (2009)
  [arXiv:0903.5407 [hep-th]].
\bibitem{Cagnazzo:2009zh}
  A.~Cagnazzo, D.~Sorokin and L.~Wulff,
  ``String instanton in AdS(4)xCP(3),''
  JHEP {\bf 1005}, 009 (2010)
  [arXiv:0911.5228 [hep-th]].
\bibitem{Chandrasekhar:2009ey} 
  B.~Chandrasekhar and B.~Panda,
  ``Brane Embeddings in AdS(4) x CP**3,''
  Int.\ J.\ Mod.\ Phys.\ A {\bf 26}, 2377 (2011)
   [arXiv:0909.3061 [hep-th]].
\bibitem{Grignani:2009ny}
  G.~Grignani, T.~Harmark, A.~Marini and M.~Orselli,
  ``New Penrose Limits and AdS/CFT,''
  JHEP {\bf 1006} (2010) 034
   [arXiv:0912.5522 [hep-th]].
  
   
\bibitem{AliAkbari:2010rs}
  M.~Ali-Akbari,
  ``A D2-brane in the Penrose limits of AdS(4)x CP(3),''
  Phys.\ Rev.\ D {\bf 82} (2010) 065027
  
 
  \bibitem{Hyun:2002wu} 
  S.~j.~Hyun and H.~j.~Shin,
  ``N=(4,4) type 2A string theory on PP wave background,''
  JHEP {\bf 0210}, 070 (2002)
  [hep-th/0208074].
  
\bibitem{Sugiyama:2002tf}
  K.~Sugiyama and K.~Yoshida,
  ``Type IIA string and matrix string on PP wave,''
  Nucl.\ Phys.\ B {\bf 644} (2002) 128
    [hep-th/0208029]
  
 
    
  
\bibitem{Alishahiha:2000qf} 
  M.~Alishahiha, H.~Ita and Y.~Oz,
  ``Graviton scattering on D6-branes with B fields,''
  JHEP {\bf 0006}, 002 (2000)
  [hep-th/0004011].
  
 
 \bibitem{Hassan:1999bv} 
  S.~F.~Hassan,
  ``T duality, space-time spinors and RR fields in curved backgrounds,''
  Nucl.\ Phys.\ B {\bf 568}, 145 (2000)
  [hep-th/9907152].
  \bibitem{Cvetic:2002si} 
  M.~Cvetic, H.~Lu and C.~N.~Pope,
  ``M theory p p waves, Penrose limits and supernumerary supersymmetries,''
  Nucl.\ Phys.\ B {\bf 644}, 65 (2002)
  [hep-th/0203229].
  \bibitem{SING}
H. Singh,  ``M5-branes with $3/8$ Supersymmetry in PP-wave background,''
[hep-th/0205020].
  
  
\end{thebibliography}
\end{document}